\documentclass[pre,twocolumn,showpacs]{revtex4}
\usepackage{epsfig}
\usepackage{amsmath}
\usepackage{amssymb}

\begin{document}

\title{Predictability of the Appearance of Anomalous Waves at Sufficiently Small Benjamin-Feir Indices}
\author{V. P. Ruban}
\email{ruban@itp.ac.ru}
\affiliation{Landau Institute for Theoretical Physics RAS, Moscow, Russia} 

\date{\today}

\begin{abstract}
The numerical simulation of the nonlinear dynamics of random sea waves at moderately small Benjamin-Feir 
indices and its comparison with the linear dynamics (at the coincidence of spatial Fourier harmonics near 
a spectral peak at a certain time $t_p$) indicate that the appearance of a rogue wave can be predicted in
advance. If the linear approximation shows the presence of a sufficiently extensive and/or high group of waves
in the near future after $t_p$, an anomalous wave is almost necessarily formed in the nonlinear model. 
The interval of reliable forecasting covers several hundred wave periods, which can be quite sufficient 
in practice for, e.g., avoiding the meeting of a ship with a giant wave.
\end{abstract}

%47.35.Bb  Gravity waves
%92.10.Hm  Ocean waves and oscillations

\maketitle

Anomalous waves or rogue waves (freak waves, giant waves) are of both applied and theoretical 
interest. They are rare individual extremely high waves among moderate waves. This subject is actively 
studied (see, e.g., reviews [1-3], special issues of journals [4, 5], and numerous references therein). 
Extremely high and steep waves with heights reaching 20-30 m at a length of 200-250 m are dangerous even 
for large ships and platforms [6, 7]. One of the characteristic properties of a rogue wave is its sudden 
appearance: ``a giant wave appears out of nowhere and disappears without any trace''. For this reason, 
it seems hopeless to predict an anomalous wave at least two hundred wave periods before its appearance 
in order to take measures. This is generally the case [8]. However, there is a widespread natural regime of seas where the
place and time of the appearance of a rogue wave can be simply and efficiently forecasted. This regime is
studied in this work. It should be emphasized that here we simulate  essentially three-dimensional 
flows of a liquid with a two-dimensional free surface, which cannot be analyzed with the methods 
and results reported in [9-15], where similar problems were studied for planar flows primarily within the 
focusing one-dimensional nonlinear Schr\"odinger equation. 

Let the main energy-carrying part of the random wave spectrum be concentrated near the wave vector ${\bf k}_0$. 
We shall speak about  sea states with moderately low Benjamin-Feir indices,  
$I_{\rm BF}\sim\tilde\nu k_0\tilde A\lesssim 1$, where $\tilde\nu$ is the average number of waves in a group, 
$k_0$ is the characteristic wavenumber, and $\tilde A$ is the typical amplitude of a wave. 
This dimensionless quantity appears naturally when describing a quasimonochromatic wave within the nonlinear 
Schr\"odinger equation [16, 17] as a parameter characterizing the presence of
nonlinear coherent structures. The case under consideration implies that such structures are almost completely absent. 
The situation where $\tilde\nu=3\dots 5$ and the vertical standard deviation of the (two-dimensional)
free surface is  $\sigma=(0.005\dots 0.007)\lambda_0$, where $\lambda_0=2\pi/k_0$
is the typical wavelength, will be considered below. It can be accepted that $\tilde A =2\sigma$; 
in this case, $I_{\rm BF}=0.3\dots 0.5$. Strictly speaking, wave groups on the two-dimensional surface are 
characterized by additional important parameters, in particular, by the characteristic area
occupied by a group. This parameter depends on the length of crests. It is assumed that the length of wave
crests is $\tilde l_{\rm cr}\sim \tilde\nu\lambda_0$ ; i.e., motion of the liquid is assumed
sufficiently three-dimensional. This wave regime is noticeably distinguished in low nonlinearity from
quasi-two-dimensional (long-crested) waves [18-22] and, the more so, from plane waves [23-26], as well as
from long-correlated three-dimensional wave fields with large Benjamin–Feir indices [27-31]. In addition, 
the action of wind on the main part of the wave state, large-scale inhomogeneous flows, vorticity of
the liquid, etc., are neglected.

The dominant dynamic factor in such random
fields is the linear dispersion with the law $\omega_{\bf k}=(gk)^{1/2}$,
where $g$ is the gravitational acceleration (waves on deep water are considered). Nonlinearity generally
plays a secondary role except for rare events where dispersion accidentally begins to form a sufficiently
extensive and/or high group of waves with the local index $\nu k_0 A\gtrsim 1$. For example, two moderate wave
groups can collide with each other. In this case, nonlinearity is strongly manifested and significantly distorts 
the two-dimensional plot of the envelope of such a group as compared to the linear theory, making it
narrower in the direction of propagation of waves and relatively longer in the transverse direction [32-34].
As a result, this leads to formation of a single, extremely nonlinear wave with a sharp, sometimes
even breaking crest (or two or three closely located anomalous waves when the orientation of the elongated 
envelope of the group is inclined; this configuration is a possible reason for the term ``three sisters'' in
sea folklore).

The above scenario implies that the vertical deviation of the free surface until the appearance of a large
group should be satisfactorily described by the simple formulas
\begin{equation}
\eta({\bf r},t)\approx \mbox{Re }A({\bf r},t), \quad 
A({\bf r},t)=\sum_{\bf k}\alpha_{\bf k}e^{i({\bf k r}-\omega_{\bf k}t)},
\end{equation}
and the Fourier components $\alpha_{\bf k}(t)$ at ${\bf k}$ values near the
spectral peak are slowly varying functions of time such that their variations hardly affect the wave dynamics
during several hundred wave periods $T_0=2\pi/\omega_0$ (although dispersion noticeably transforms the picture
of the wave field in this time interval). Thus, a reasonable approximation is
\begin{equation}
\alpha_{\bf k}(t)\approx \alpha_{\bf k}(t_p), \quad \mbox{if} 
\quad |t-t_p|\lesssim T_0/(k_0^2\sigma^2).
\end{equation}
This estimate follows from a known formula for the nonlinear frequency shift of a quasimonochromatic
wave $\delta\omega_{\rm nonlin}=\omega_0k_0^2|A|^2/2$. Formulas (1) and (2) correspond to the so-called 
second-order theory (see reviews [1-3] and references therein), where resonant
four-wave processes are neglected, whereas nonresonant three-wave interactions are taken into account in
the form of the second harmonic.

\begin{figure}
\begin{center}
   \epsfig{file=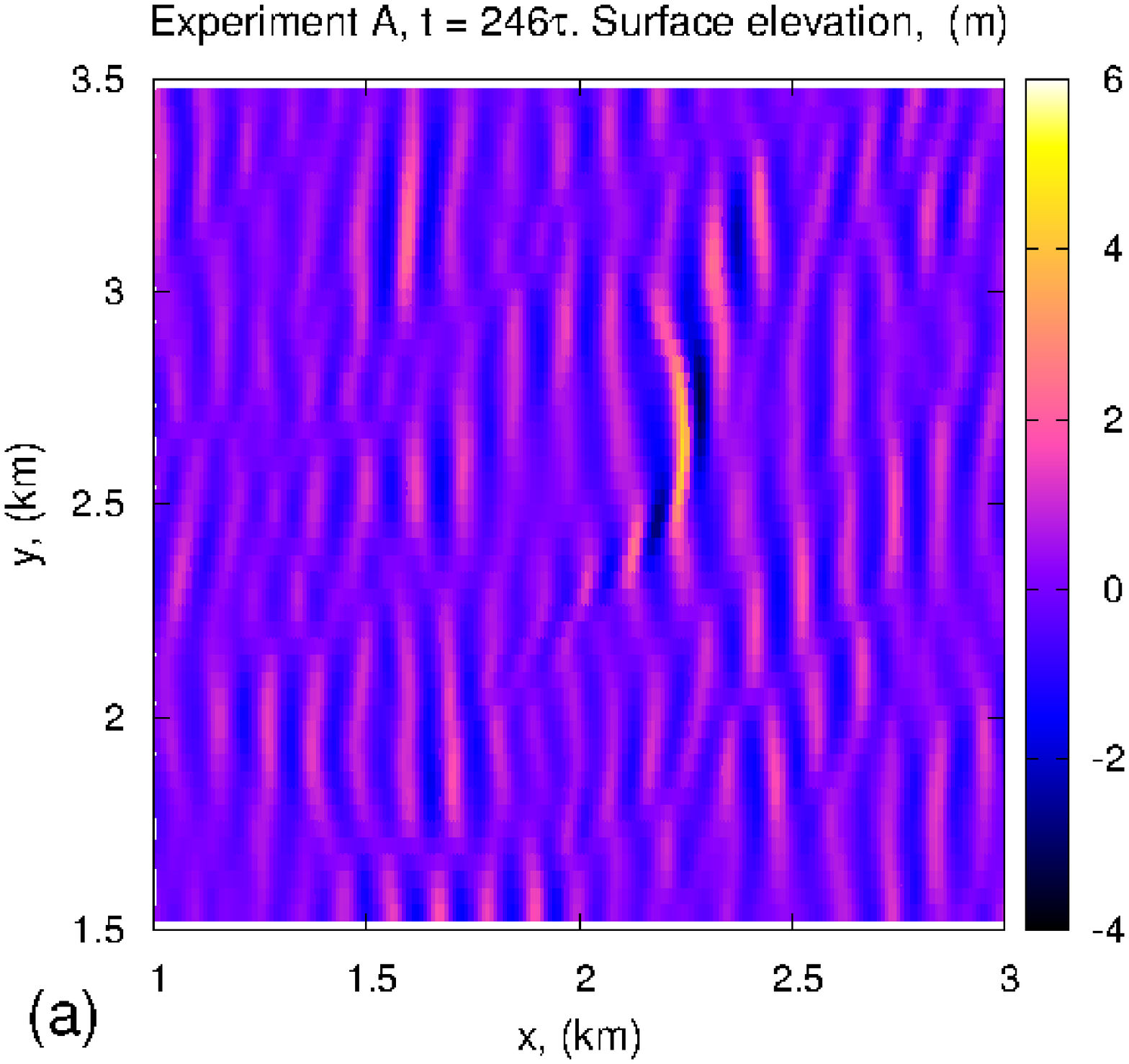, width=75mm}\\
\vspace{3mm}
   \epsfig{file=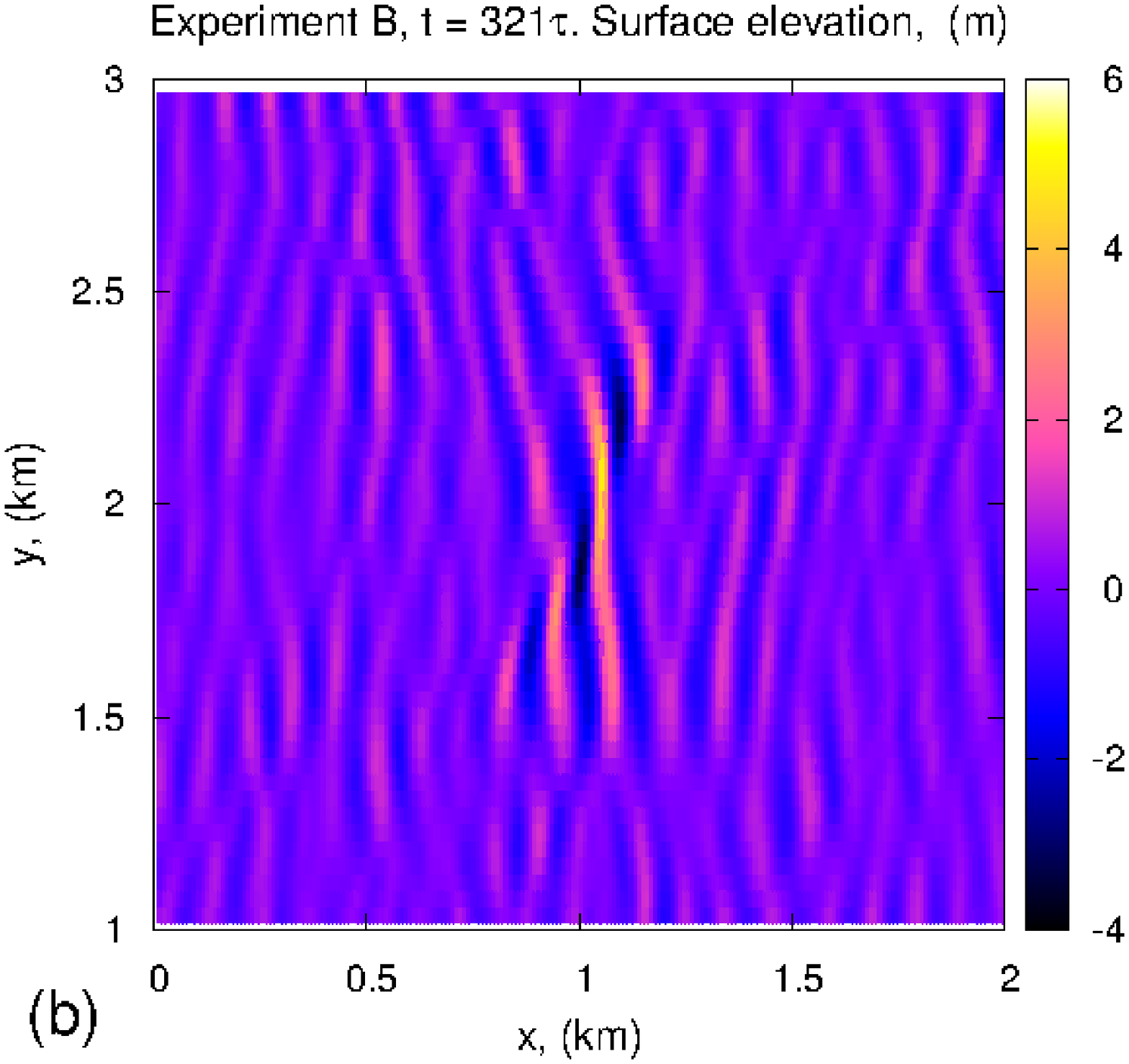, width=75mm}\\
\vspace{3mm}
   \epsfig{file=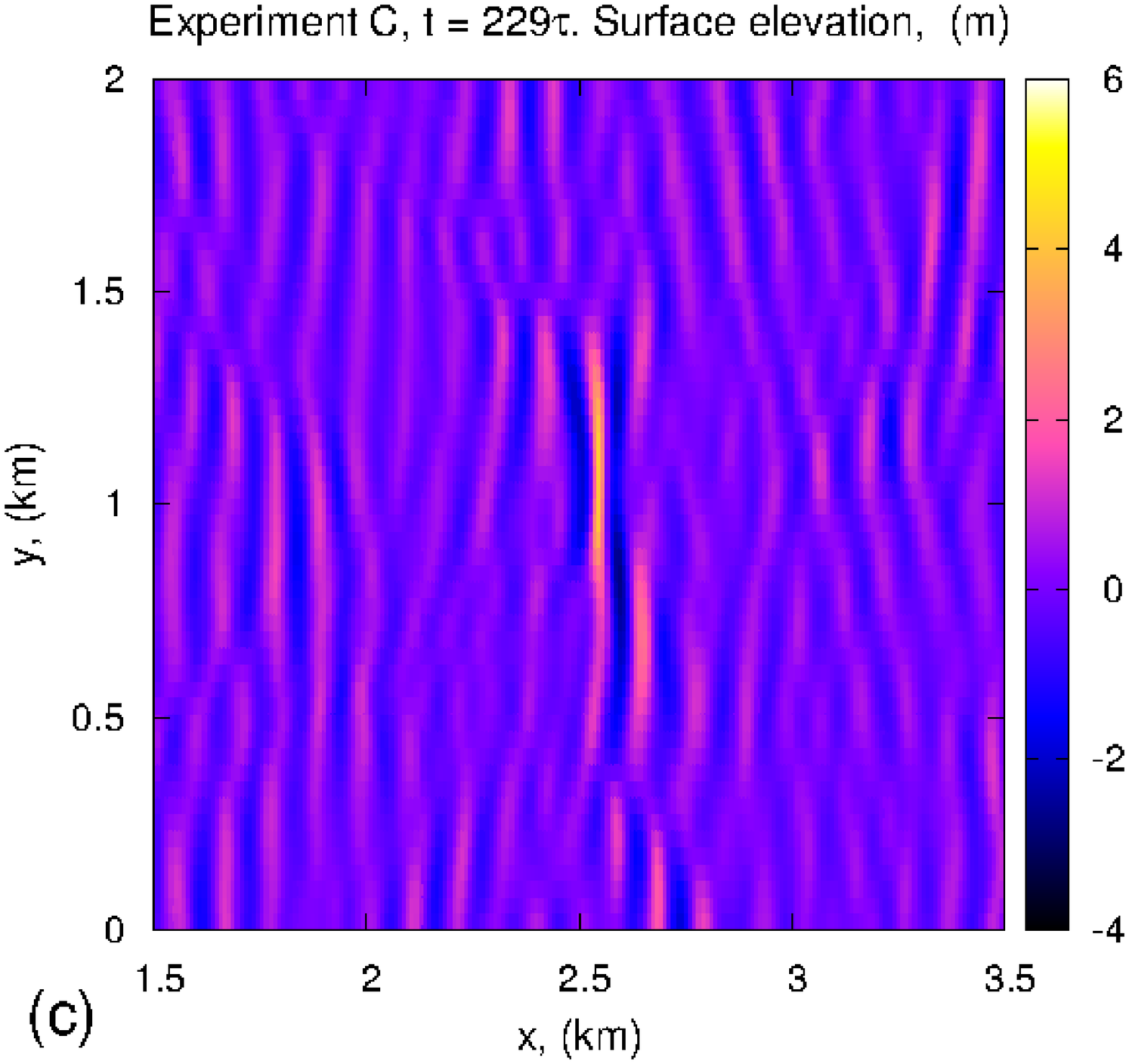, width=75mm}
\end{center}
\caption{Anomalous waves formed in three numerical experiments.}
\label{RW-maps} 
\end{figure}
\begin{figure}
\begin{center}
   \epsfig{file=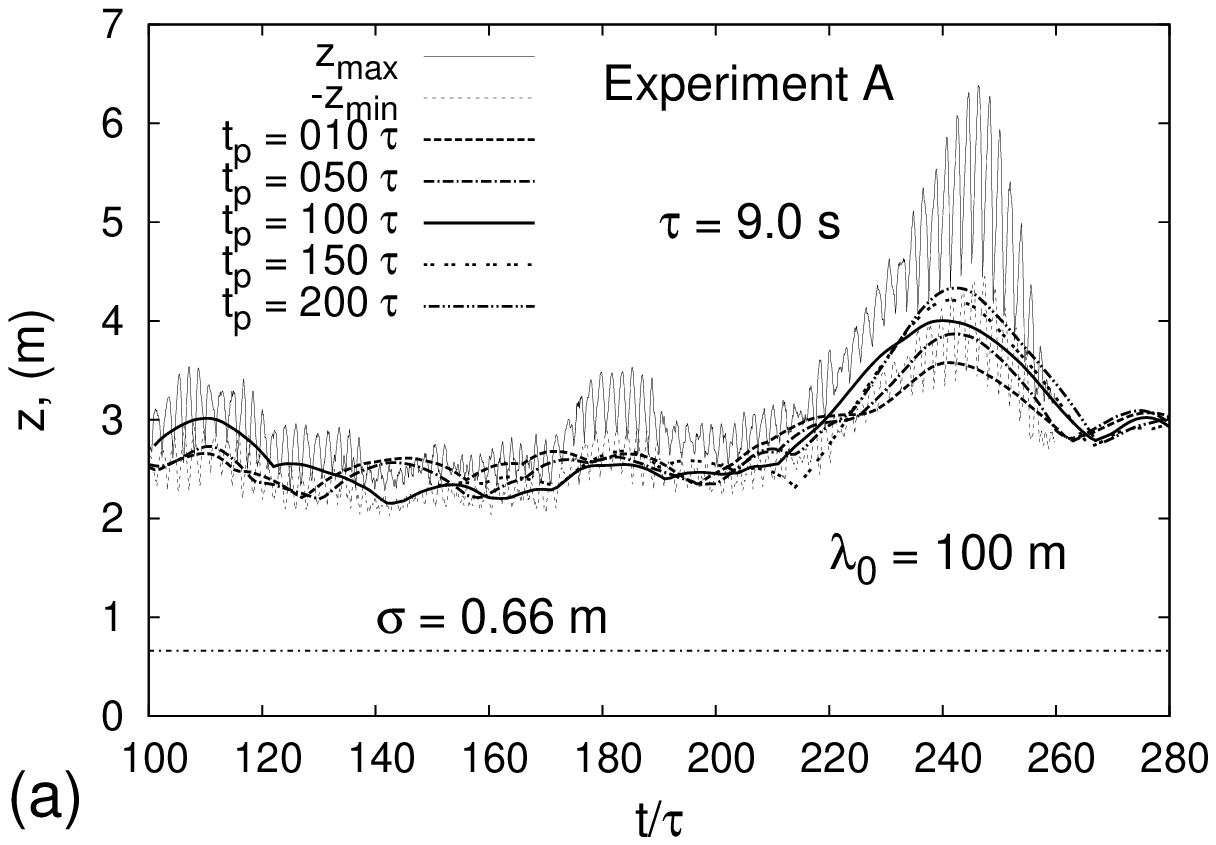, width=80mm}\\
\vspace{3mm}
   \epsfig{file=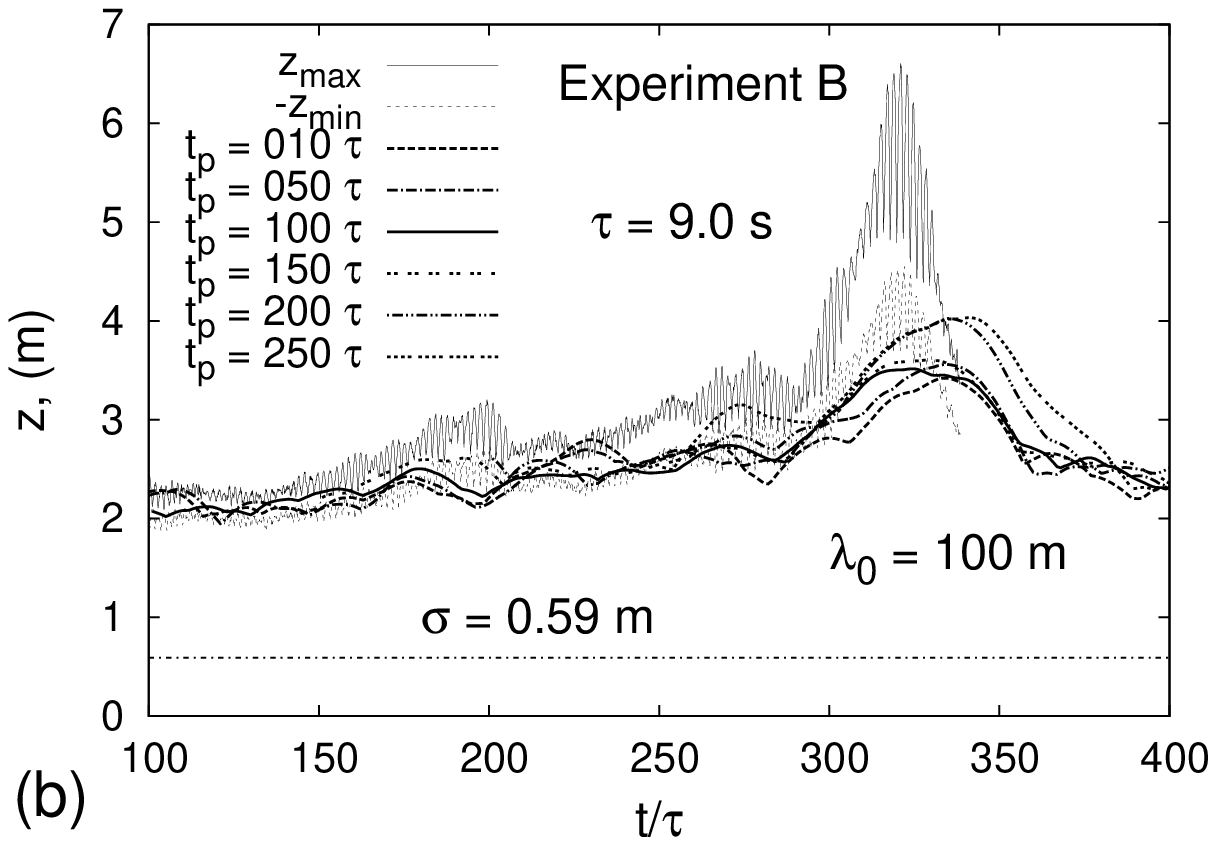, width=80mm}\\
\vspace{3mm}
   \epsfig{file=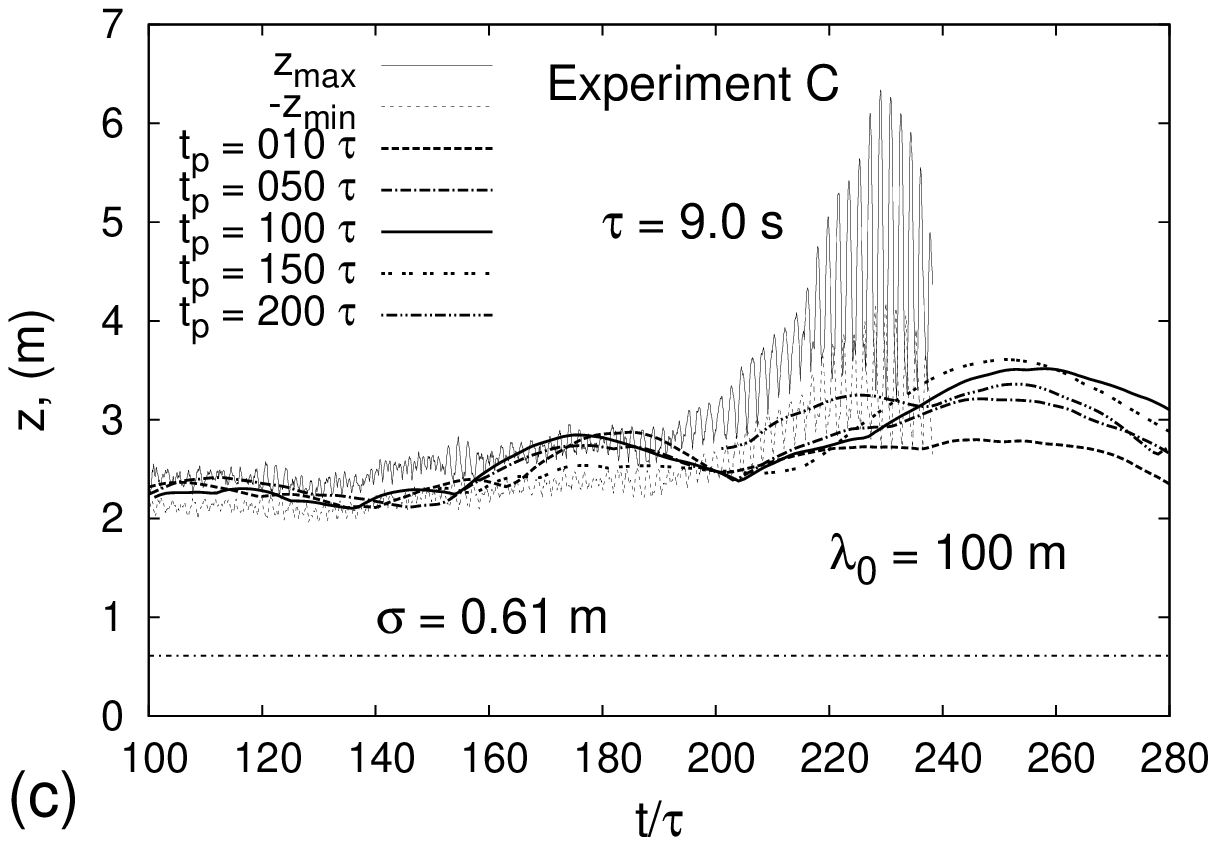, width=80mm}
\end{center}
\caption{Extremal deviations of the free surface in comparison with linear forecasts for different  $t_p$.}
\label{predictions} 
\end{figure}
\begin{figure}
\begin{center}
   \epsfig{file=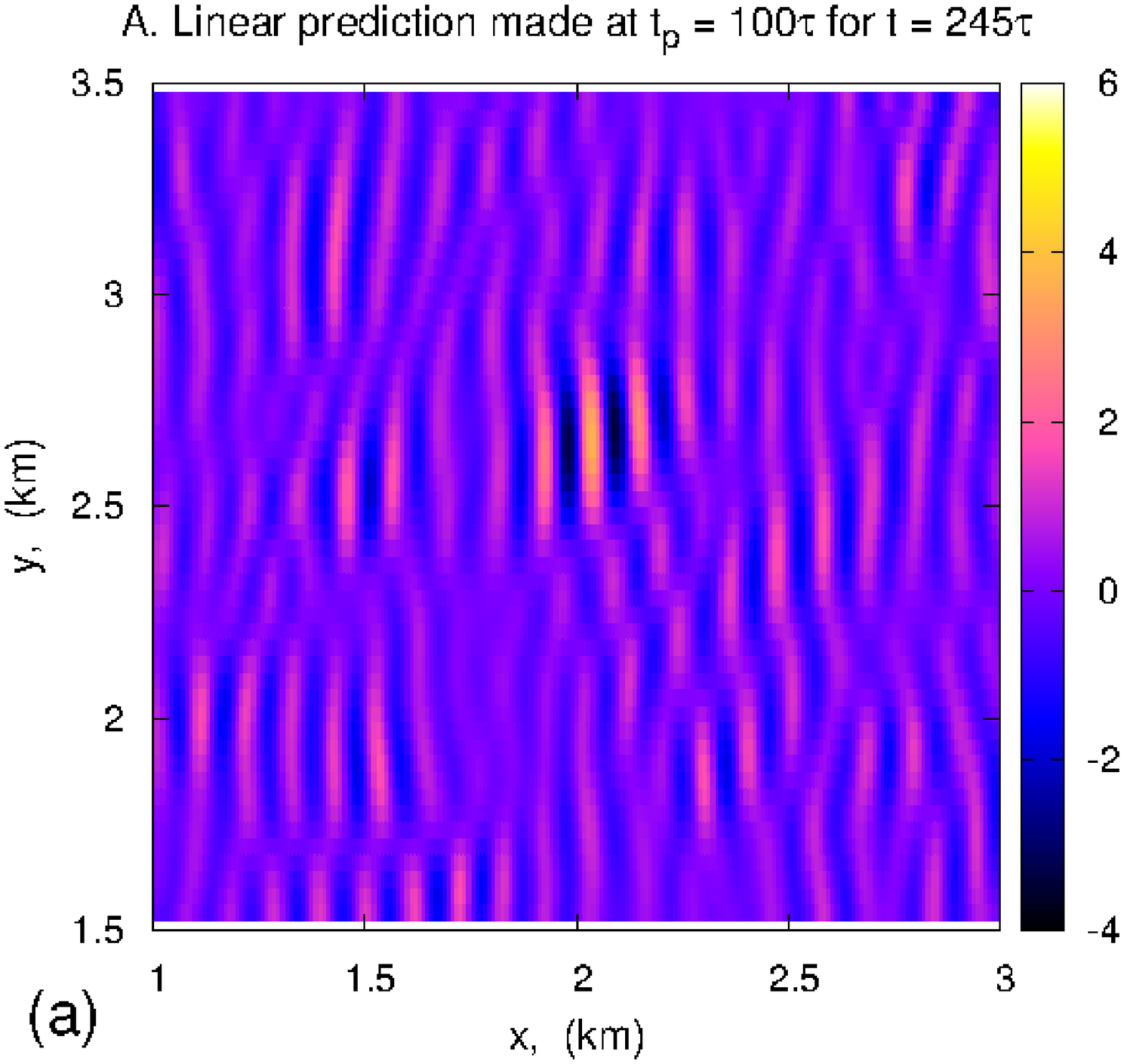, width=75mm}\\
\vspace{3mm}
   \epsfig{file=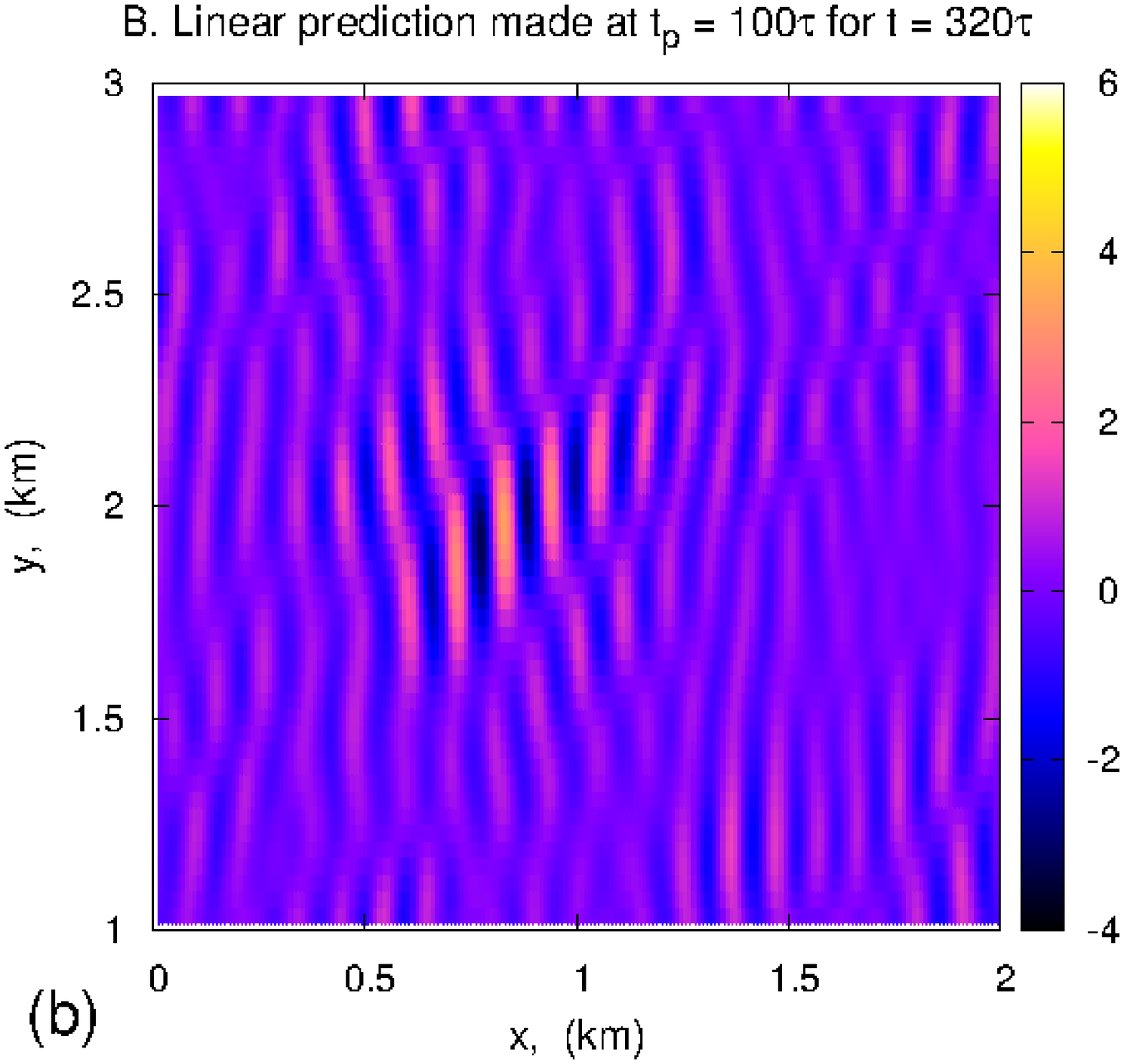, width=75mm}\\
\vspace{3mm}
   \epsfig{file=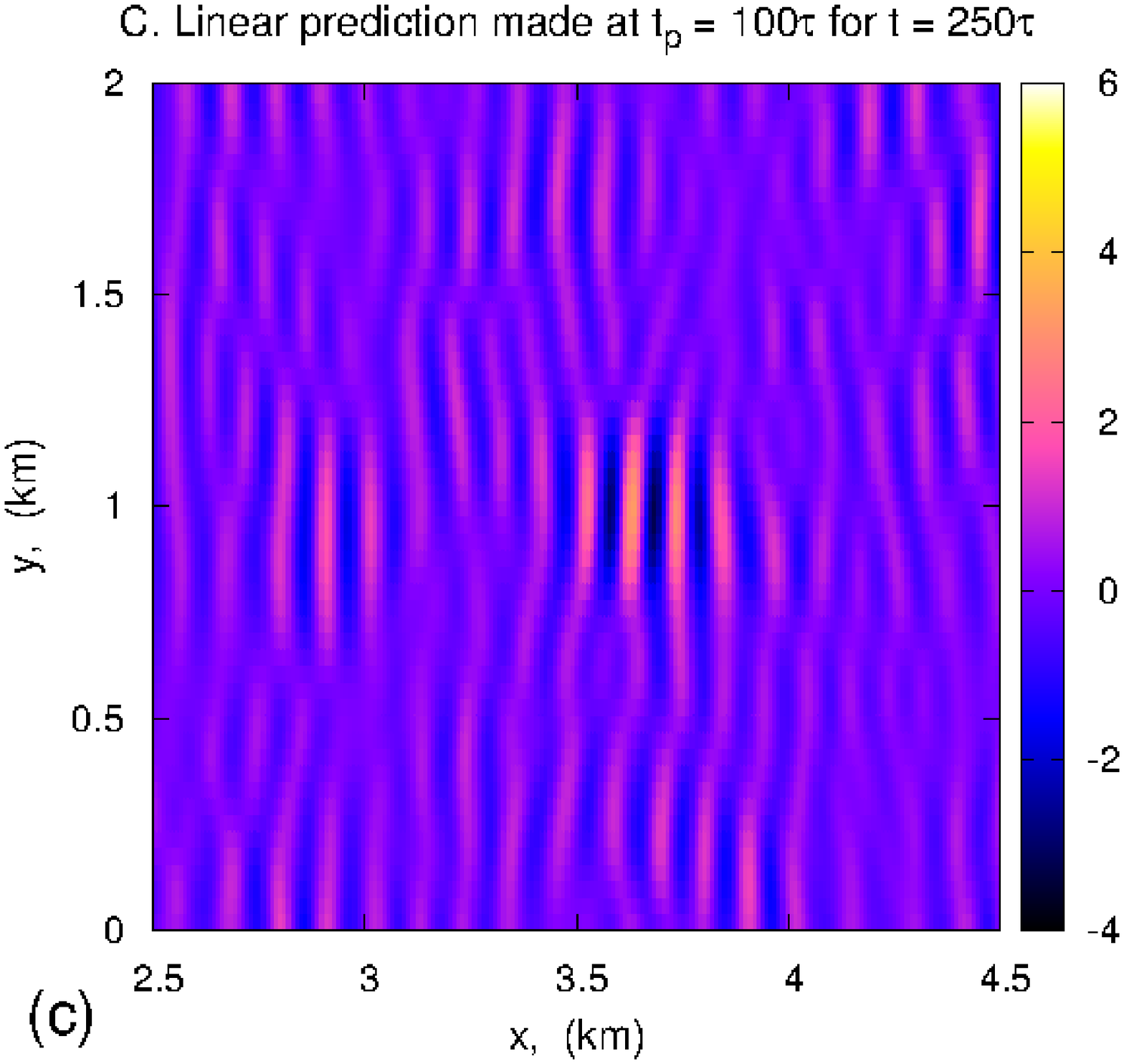, width=75mm}
\end{center}
\caption{ Linear forecasts indicate the formation of sufficiently high groups of waves.}
\label{predicted_groups-maps} 
\end{figure}

\begin{figure}
\begin{center}
   \epsfig{file=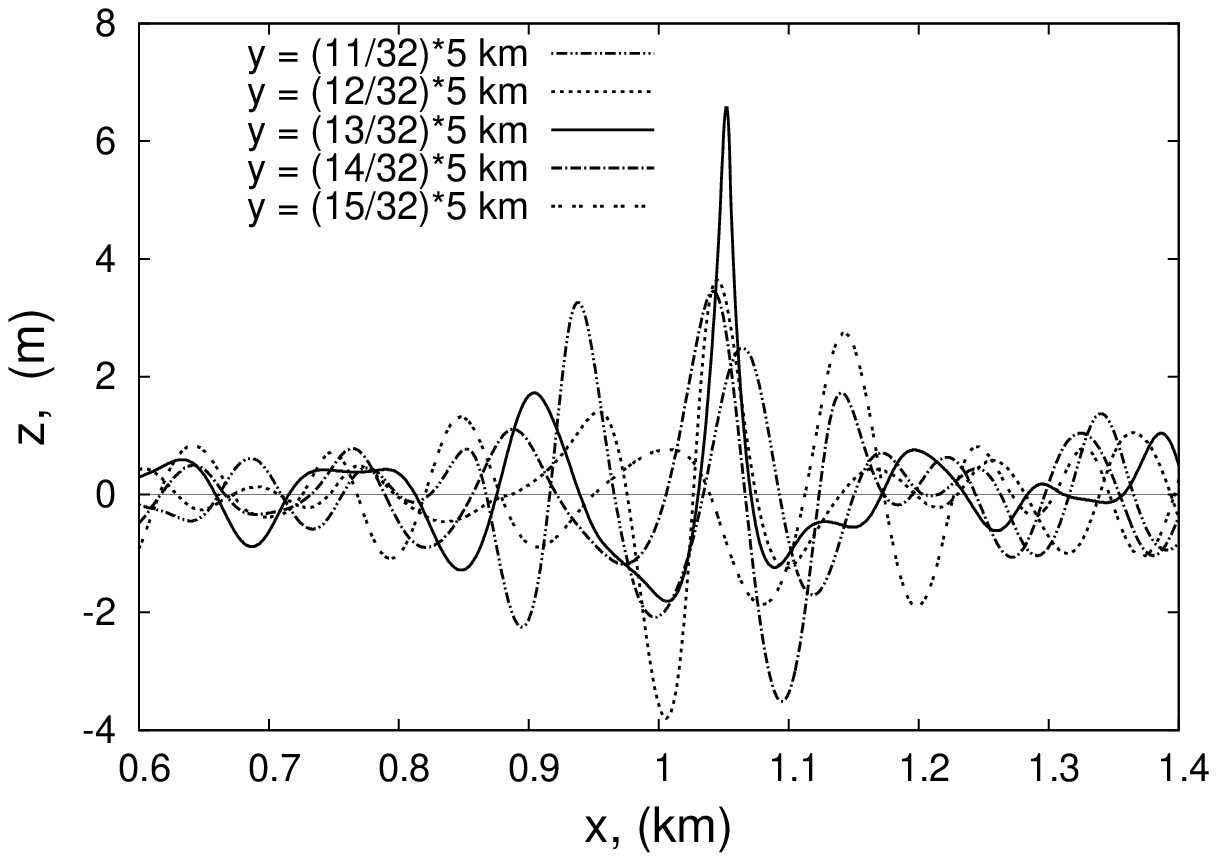, width=85mm}\\
\end{center}
\caption{Several wave profiles from Fig. 1b.}
\label{wave_profiles} 
\end{figure}

It is important that Eqs. (1) and (2) allow forecasting the appearance or absence of a rogue wave at $t>t_p$
if the spatial spectrum $\eta_{\bf k}$  of the vertical deviation of the free surface is measured at the time $t_p$ in a 
sufficiently large region of the sea. A map of the sea surface can be obtained from satellites and be digitized; then,
the fast Fourier transform can be performed. The forecasting procedure is as follows.

First, the complex conjugate harmonics of the real function $\eta({\bf r},t_p)$ near $(-{\bf k}_0)$ should be rejected, 
and $\alpha_{\bf k}=2\eta_{\bf k}$ should be set at $(+{\bf k}_0)$. It is also desirable to
filter high wavenumbers, retaining only the energy-carrying vicinity of the spectral peak. This will allow
using a large spatial resolution and accelerating computer calculations.

Second, it is necessary to perform a linear forecast of the sea state by Eqs. (1) and (2) with a time step of
about (1–2)$T_0$. Such a forecast on a PC takes about a minute. In particular, the time dependence of the maximum 
$|A|$ value in the region under consideration should be plotted.

Further, the dependence $|A|_{\rm max}(t)$ should be analyzed. The possible appearance of a maximum on the plot at 
$t_*(t_p)$ soon after $t_p$ with the height exceeding, say, $6\sigma$ is a serious indication of the nucleation of an
anomalous wave.

Additional information can be obtained from the map of the sea surface at $t=t_*(t_p)$ plotted according to
linear forecasting. The place of the appearance of the rogue wave, as well as the local Benjamin-Feir index,
can be estimated on the map. The larger the local Benjamin-Feir index, the larger the expected nonlinear
increase in the height of the wave as compared to the linear prediction.

Three numerical examples presented below demonstrate the efficiency of the above method for
forecasting anomalous waves. The wave dynamics was calculated within the completely nonlinear weakly
three-dimensional model [35, 36] in a $2\pi\times 2\pi$ square
with periodic boundary conditions and the dimensionless gravitational acceleration $\tilde g =1$, where waves
propagate on average along the x axis. The characteristic wavenumber was chosen to be 50. In the recalculation 
to a square with a side of 5 km and  $g=$9.81 m/s$^2$,
the dimensionless time unit corresponds to $\tau=[5000/(9.81\cdot 2\pi)]^{1/2}=9.01$ s. 
In this case, the wave-length is  $\lambda_0 = $ 100 m and the period is $T_0\approx 8$ s.

In all three numerical experiments, quasirandom
initial data for nonlinear simulation were chosen by
means of a special procedure (which is not described
here) such that an anomalous wave is formed in the
system in a time of (250-350) $T_0$. The corresponding
“portraits” of these rogue waves near the times of their
highest elevation are shown in Fig. 1. All three waves
undergo breather oscillations because of the difference
between the phase and group velocities, particularly,
the ``straight'' wave C (in contrast to ``inclined'' waves
A and B, where crests and troughs seemingly move
along the inclined elongated envelope from its ``end''
to the ``beginning''; this motion can be visually
observed as the lateral propagation of the anomalous
wave).

Figure 2 shows the results of nonlinear calculations
of the height of the highest crest and the depth of the
deepest trough in comparison to linear forecasts made
at different times $t_p$. As a whole, the closer the time $t_p$
to the time of rise of the rogue wave, the clearer the
prediction of the formation of a high group by the linear theory. 
Figure 3 shows examples of such “linear”
groups. Differences from anomalous waves in Fig. 1
are very significant. It is seen that even the longest
group in case B ``has been transformed'' by nonlinearity in a
single, very high wave in spite of a somewhat smaller
average Benjamin-Feir index. It is remarkable that a
large linear group appears owing to the collision of two
wave groups (not shown in the figures), which confirms the assumption made in [34] that this 
mechanism is topical. In Fig. 1b, two deep troughs are
located slantwise on two opposite sides of the high
crest near it and two less high crest are located further.
In general, such a configuration corresponds to three sisters. The
height of the crest in this case is more than an order of
magnitude (!) larger than the standard deviation and
the crest itself is very sharp, as is seen in Fig. 4. It is
clear that the second-order theory is inapplicable to
the description of such really anomalous waves. Furthermore, the rogue wave in case C rises noticeably
earlier in time (and closer along the trajectory of motion)
than that predicted by the linear model. Such nonlinear effects should be taken into account when 
forecasting the time and place of the appearance of a large wave.

It is also noteworthy that, in very rare cases, a long-term forecast can predict the appearance of a large
group in a time of, e.g., $(200\dots 300) T_0$ after $t^{(1)}_p$,
but an anomalous wave does not appear in reality. However, later forecasts with
$t^{(2)}_p>t^{(1)}_p$, $t^{(3)}_p>t^{(2)}_p$, etc., approaching the forecasting time will show gradually 
smaller linear groups at the indicated time. For this reason, the method is sufficiently reliable.

According to the linear estimate of the frequency of
appearance of extreme waves, a high group exceeding
$6\sigma$ appears on average once in  $(4\dots 8)\times 10^4 T_0$ on an
area of $(50)^2\lambda_0^2$. This means that a giant wave at the taken
parameters appears on an area of, e.g., $50\times 50$ km$^2$
once in several hours. Obviously, less anomalous but
quite high waves in the form of groups of two or three
waves appear much more frequently. They correspond
to smaller linear groups in the forecast (they are not
shown in the figures).

To summarize, the fundamental possibility of
advanced prediction of anomalous waves in sea states
with sufficiently small Benjamin-Feir Indices within
the linear model has been demonstrated. The problem
of possible implementation and applied usefulness of
such an approach requires additional studies and discussions. 
In particular, it is still unclear how the interaction of waves
with wind and other factors disregarded in our model can change the quality of the
forecast. For this reason, details of the method such as
the choice of the sizes of the forecasting region and
boundary conditions at the Fourier transformation, as
well as the possibility of estimating the parameters of
anomalous waves in the nonlinear stage by using the
Gaussian variation model considered in [33] or by
modifying the forecasting equation by introducing
local nonlinearity such as the nonlinear Schrödinger
equation and using the split-step method, were not
discussed.

\end{document}